\documentclass[lettersize,journal]{IEEEtran}
\usepackage{amsmath,amsfonts}
\usepackage{algorithmic}
\usepackage{xspace}
\usepackage{algorithm}
\usepackage{array}
\usepackage[caption=false,font=normalsize,labelfont=sf,textfont=sf]{subfig}
\usepackage{textcomp}
\usepackage{stfloats}
\usepackage{xcolor}
\usepackage{url}
\usepackage{verbatim}
\usepackage{graphicx}
\usepackage{multirow}
\usepackage{tabularx}
\usepackage{cite}
\usepackage{amsthm}
\usepackage{booktabs}
\usepackage{multirow}
\newtheorem{definition}{Definition}
 
\newcommand{\method}{\texttt{NoD-DGMT}}

\providecommand{\siglip}{\mbox{\sc{SigLIP}}\xspace}
\providecommand{\dino}{\mbox{\sc{DinoV2}}\xspace}
\providecommand{\ObjectNav}{\mbox{\sc{ObjNav}}\xspace}
\providecommand{\bench}{\mbox{\sc{Chores}}\xspace}

\providecommand{\benchnav}{\mbox{\sc{ChoresNav}}\xspace}
\providecommand{\fifteennodash}{\mbox{$\mathbb{S}$}\xspace}
\providecommand{\fifteen}{\mbox{-\fifteennodash}\xspace}
\newcolumntype{C}{>{\centering\arraybackslash}X}

\begin{document}

\title{Detecting Non-Optimal Decisions of Embodied Agents via Diversity-Guided Metamorphic Testing}

\author{Wenzhao Wu, Yahui Tang,  Mingfei Cheng, Wenbing Tang, \\Yuan Zhou,~\IEEEmembership{Member,~IEEE}, and Yang Liu,~\IEEEmembership{Senior Member,~IEEE}
\thanks{W. Wu is with the National Supercomputing Center, Wuxi, Jiangsu 214072,
P. R. China (E-mail: wumz13@tsinghua.org.cn).}
\thanks{Y. Tang is with the School of Computer, Chongqing University of Posts and Telecommunications, Chongqing, 400065, P. R. China (E-mail: 
tangyh@cqupt.edu.cn).}
\thanks{M. Cheng is with the School of Computing and Information Systems,  Singapore Management University, Singapore 188065 (E-mail: 
mfcheng.2022@smu.edu.sg).}
\thanks{W. Tang is with the College of Information Engineering,  Northwest A\&F University, 
Yangling, Shaanxi 712100, P. R. China (E-mail: 
wenbingtang@nwafu.edu.cn).}
\thanks{Y. Zhou is with the School of Computer Science and Technology, Zhejiang Sci-Tech University, Hangzhou, Zhejiang 310018, P. R. China (E-mail: yuanzhou@zstu.edu.cn).}
\thanks{Y. Liu is with the College of Computing and Data Science, Nanyang Technological University, Singapore 639798 (E-mail: 
yangliu@ntu.edu.sg).}
}



\maketitle

\begin{abstract}
As embodied agents advance toward real-world deployment, ensuring optimal decisions becomes critical for resource-constrained applications. Current evaluation methods focus primarily on functional correctness, overlooking the non-functional optimality of generated plans. This gap can lead to significant performance degradation and resource waste. We identify and formalize the problem of Non-optimal Decisions (NoDs), where agents complete tasks successfully but inefficiently.
We present \method, a systematic framework for detecting NoDs in embodied agent task planning via diversity-guided metamorphic testing. Our key insight is that optimal planners should exhibit invariant behavioral properties under specific transformations. We design four novel metamorphic relations capturing fundamental optimality properties: position detour suboptimality, action optimality completeness, condition refinement monotonicity, and scene perturbation invariance. To maximize detection efficiency, we introduce a diversity-guided selection strategy that actively selects test cases exploring different violation categories, avoiding redundant evaluations while ensuring comprehensive diversity coverage.
Extensive experiments on the AI2-THOR simulator with four state-of-the-art planning models demonstrate that \method~achieves violation detection rates of 31.9\% on average, with our diversity-guided filter improving rates by 4.3\% and diversity scores by 3.3 on average. \method~significantly outperforms six baseline methods, with 16.8\% relative improvement over the best baseline, and demonstrates consistent superiority across different model architectures and task complexities.
\end{abstract}

\begin{IEEEkeywords}
Decision Optimality,  Embodied Agents, 
Metamorphic Testing, Task Planning, Violation Diversity
\end{IEEEkeywords}

\section{Introduction}

While foundation models, such as Large Language Models (LLMs) and Vision-Language-Action (VLA) models, have demonstrated impressive capabilities in natural language understanding, multimodal reasoning, and content generation, they remain fundamentally disembodied and lack the ability to directly interact with the physical world~\cite{yang2024embodied}. This trend has led to the emergence of embodied AI~\cite{liu2025embodied,mon2025embodied,feng2025multi}, where embodied agents are equipped with physical forms such as robots, enabling them to actively perceive, learn from, and interact with the physical environment. Consequently, embodied agents are widely regarded as a pivotal step toward achieving Artificial General Intelligence (AGI), given their potential in various real-world applications~\cite{ge2024behavior,fan2025embodied,liu2025survey,zhou2024embodied,huang2024drivlme}, such as household services, industrial manufacturing, healthcare, and autonomous driving.

In embodied agents, task planning constitutes a fundamental capability that encompasses the decision-making processes required to decompose complex task objectives into executable action sequences~\cite{zhang2024plan,chang2024partnr}. This involves sophisticated multi-dimensional reasoning, wherein agents must interpret natural language instructions, accurately assess environmental states and constraints, and effectively coordinate interdependencies among sequential subtasks~\cite{wang2025world}. The resulting action plans must be both computationally feasible and contextually appropriate for successful task execution.
Recently, numerous studies have explored leveraging LLMs for embodied task planning, demonstrating promising performance and enhanced adaptability~\cite{choi2024lota,zhang2024plan,lei2025stma,li2024embodied}. As these agents become increasingly deployed in safety-critical and resource-sensitive scenarios, ensuring the quality and reliability of their decisions becomes paramount.

However, existing evaluation methods for embodied agents primarily focus on basic functional requirements, such as task success rates~\cite{ehsani2024spoc,hu2024flare,zhang2024lamma}, trajectory length~\cite{zhong2025p3nav,chattopadhyay2021robustnav}, and execution time~\cite{zhang2024plan}.
Although these functional metrics provide valuable insights into task completion capabilities, they fail to capture crucial non-functional requirements essential for real-world deployment.
\textit{One important yet underexplored aspect is optimality, which is the ability to make decisions that minimize resource consumption and maximize efficiency through the most effective means possible.} 
While Non-optimal Decisions (NoDs) do not compromise task completion, they can result in substantial cost increases and resource inefficiencies in practical scenarios. 
For instance, suboptimal planning in household service robots may lead to excessive energy consumption and user frustration due to prolonged operation times.
Therefore, developing a comprehensive framework to systematically detect and evaluate NoDs is imperative for ensuring practical utility and cost-effectiveness in resource-constrained deployment scenarios.

Systematically detecting and evaluating NoDs in embodied agents presents several significant challenges. 
First, unlike functional testing with explicit criteria (e.g., task completion rates), optimality assessment lacks clear oracles for determining the ``optimal'' action sequence across diverse scenarios and contexts.
Second, ensuring comprehensive NoD detection is challenging, as evaluation processes 
may repeatedly identify similar suboptimal patterns while overlooking other violation categories, thereby limiting coverage comprehensiveness.
Third, while functional testing relies on direct input-output comparisons, optimality evaluation requires establishing meaningful comparison criteria and appropriate metrics to assess the relative performance of different decision sequences.


To address these challenges, we propose \method, a novel Diversity-Guided Metamorphic Testing (MT) framework designed to detect NoDs in embodied agent task planning.
Specifically, we first identify and formalize four Metamorphic Relations (MRs) that capture the invariant properties of optimal decisions, enabling optimality assessment without explicit oracles. Second, we define diversity metrics and propose a diversity-guided selection strategy that filters and prioritizes test cases with greater variations to ensure comprehensive evaluation coverage across different types of potential NoD patterns. 
Finally, we develop an MT-based detection framework that leverages the identified MRs and diverse test cases to automatically evaluate decision optimality and quantify violation severity.

We conduct comprehensive experiments to evaluate the effectiveness and efficiency of \method~using the AI2-THOR simulator~\cite{kolve2017ai2}.
First, we demonstrate \method's effectiveness across varying task complexities with four different planners, successfully detecting NoDs with violation rates ranging from 18.7\% to 67.2\%.
Second, ablation studies validate our diversity-guided filter, showing it enhances detection efficiency by 4.3\% while improving diversity coverage.
Finally, comparison with four baseline methods demonstrates \method's superiority, achieving 31.9\% average detection rate while baseline methods achieve only 5.7\%.


The main contributions of this paper are as follows.

\begin{itemize}
\item We propose \method, the first MT-based framework to evaluate the optimality of embodied agent decisions.
\item We introduce four innovative MRs and a diversity-guided test case filter to systematically and efficiently discover diverse NoDs.
\item We conduct extensive evaluations across multiple planners, various task complexities, and different testing scenarios, to demonstrate the effectiveness and efficiency of \method.
\end{itemize}

The rest of this paper is organized as follows.
Section~\ref{background} introduces the background.
Section~\ref{motivation} presents the motivation and problem statement.
The detailed procedures of \method~are provided in 
Section~\ref{approach-details}.
Experiments are conducted in Section~\ref{experiment-sec} to demonstrate the effectiveness and efficiency of our approach.
Section~\ref{related-work-sec} summarizes the related work.
Finally, Section~\ref{conclusion-sec} presents the discussion and conclusion.

\section{Background}
\label{background}

\subsection{Embodied Task Planning}

\begin{figure}[!t]
    \centering
    \includegraphics[width=0.48\textwidth]{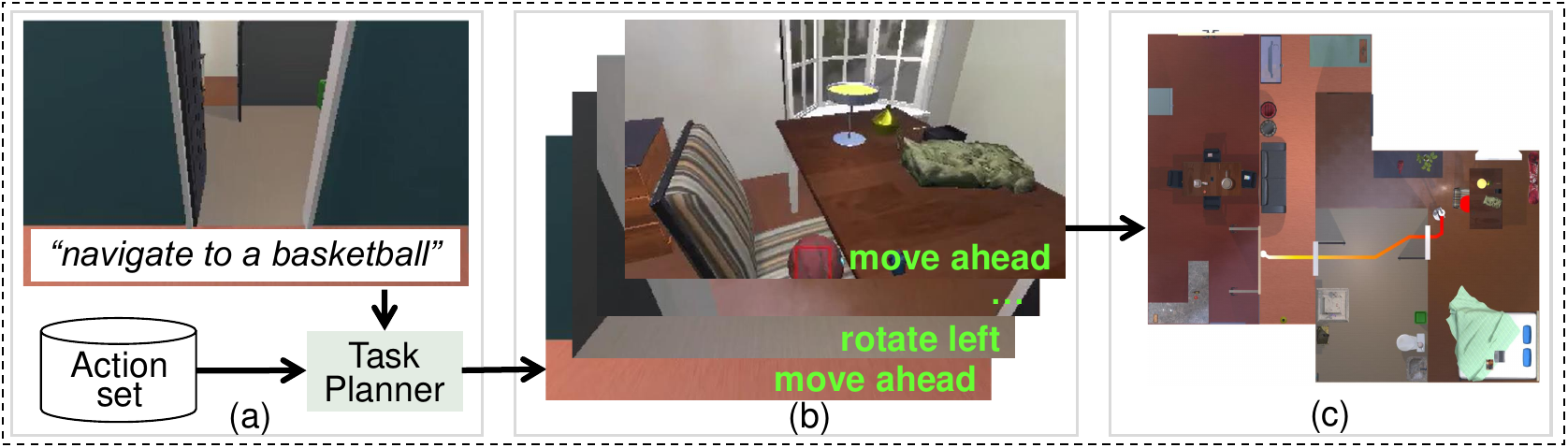} 
    \caption{The task planning process of embodied agents.}
    \label{fig:task-planning-process}
\end{figure}

Embodied task planning is a process to generate a sequence of executable actions to accomplish the given goal within a physical environment ~\cite{zhang2024plan,zhang2024lamma}.
As shown in Fig.~\ref{fig:task-planning-process}(a), the planning process begins when the task planner receives a natural language instruction~$\mathcal{L}$ from the user, such as \textit{"navigate to a basketball."}
Then, the planner  selects a sequence of actions from the predefined action set~$\mathbf{A}$
based on its perception of the 3D interactive physical environment~$E$.
The action set~$\mathbf{A}$ represents the embodied agent's capabilities, where each action~$a \in \mathbf{A}$ corresponds to an atomic operation the agent can execute (e.g.,
\textit{``\MakeUppercase{move ahead}"}, \textit{``\MakeUppercase{move back}"},  \textit{``\MakeUppercase{rotate left}"},
\textit{``\MakeUppercase{rotate right}"}, or \textit{``\MakeUppercase{move arm up}"}), with each action being coupled to a low-level motion controller.
Formally, the task planner is a policy $\Pi: E \times \mathcal{L} \rightarrow \mathbf{A}$ that maps environment states and instructions to actions. 
As shown in Fig.~\ref{fig:task-planning-process}(b), at each time step $t$, the planner selects action $a_t$ by maximizing the likelihood of completing the given instruction $\mathcal{L}$:
\begin{equation}
    a_t = \arg\max_{a \in \mathbf{A}} p(a|\mathcal{L}; a_0, \cdots, a_{t-1}; E).
\end{equation}
where $\langle a_0, \cdots, a_{t-1}\rangle$ represent the actions previously selected for time steps $\langle0, \cdots, t-1\rangle$, respectively.
A task is considered successfully completed when the given instruction $\mathcal{L}$ is satisfied at the terminal time step $\tau$:
\[
r = 
\left\{
\begin{array}{l}
1, \quad \mathrm{if}~s_{\tau} \models \mathcal{L}~ (\mathrm{instruction~satisfied}), \\
0, \quad \mathrm{otherwise}.
\end{array}
\right.
\]
where $s_{\tau}$ denotes the state of the embodied agent and the environment at time step $\tau$.
Fig.~\ref{fig:task-planning-process}(c) shows the traveled trajectory of the agent when it successfully arrives at the terminal state.

\subsection{Non-Optimal Decisions (NoDs)}

In practice, for a given instruction $\mathcal{L}$ in an environment $E$, there often exist multiple valid action sequences that can successfully complete the task. 
This paper focuses on detecting NoDs to identify and evaluate the suboptimality of planning decisions.
NoDs represent a category of decision-making issues where the generated action sequences successfully complete assigned tasks but exhibit suboptimal characteristics compared to more efficient alternatives. 
Unlike functional bugs that prevent task completion, NoDs manifest as inefficient yet functionally correct solutions that achieve the desired objectives while incurring unnecessary costs.

\begin{definition}[NoDs]
Let $\boldsymbol{\Theta}(\mathcal{L}, E)$ denote the set of all valid action sequences that successfully complete task $\mathcal{L}$ in environment $E$. Each sequence $\theta = \langle a_0, a_1, \ldots, a_{\tau}\rangle$ satisfies the functional correctness criterion where $s_{\tau} \models \mathcal{L}$. Let $C(\theta)$ be a cost function that quantifies the quality of sequence $\theta$, incorporating factors such as execution time, action count, energy consumption, or path length.
An action sequence $\theta$ generated by an embodied agent is a NoD iff:
\begin{equation}
\theta \in \boldsymbol{\Theta}(\mathcal{L}, E) \text{ s.t. } \exists \theta^\prime \in \boldsymbol{\Theta}(\mathcal{L}, E) : C(\theta^\prime) < C(\theta).
\end{equation}
\end{definition}

This definition captures the fundamental characteristic that NoDs are not functional failures but rather suboptimal solutions where more efficient alternatives exist within the same problem space.
In resource-constrained deployment scenarios, even minor inefficiencies can accumulate into significant operational costs over extended periods.
Unlike traditional software defects that cause system crashes or incorrect outputs~\cite{gao2025clozemaster,wang2025moditector,liu2024testing}, NoDs represent a class of performance issues that compromise system efficiency while maintaining functional correctness.

\subsection{Metamorphic Testing (MT)}

Unfortunately, it is hard to evaluate whether the decision $\theta$ is an optimal one due to the lack of testing oracles, i.e., we do not know the optimal plan $\theta^* = \arg\min_{\theta'} C(\theta')$.
Inspired by the success of MT in automatically detecting bugs, uncovering defects, and revealing biases in software and AI systems~\cite{zhu2024mtas,xu2024evaluating,xu2024mr,xie2024metamorphic,paltenghi2023morphq}, we adopt MT to evaluate decision optimality in embodied agents without requiring explicit optimal solutions. At the core of MT lies a set of MRs, which specify necessary properties of the system under test by defining relationships between multiple inputs and their expected outputs.
For example, consider a monotonically increasing function \( f(x) \), a simple MR is that for any two inputs where \( x_1 < x_2 \), the outputs should satisfy \( f(x_1) < f(x_2) \). The goal of MT is to construct source test cases and their corresponding follow-up test cases based on the MRs. These pairs of test cases are then executed using the target software, and the outputs of the source and follow-up test cases are compared to check whether their corresponding MRs are satisfied. If an MR is violated, we can conclude that a failure is detected. For example, if \( f(3) > f(5) \) for the test cases 
3 and
5, this discrepancy suggests an error in the function implementation, thereby indicating a failure in maintaining the monotonic property.


In this work, we apply MT to systematically detect NoDs in embodied agent task planning by formulating a set of MRs that encode the expected behavioral properties of optimal decisions. This approach enables us to determine whether a decision exhibits suboptimal behavior by checking if these MRs are violated, without requiring knowledge of the optimal decision itself. For example, one MR could specify that removing obstacles from an agent's path should not increase the number of steps in the resulting plan. If a planner violates this property, it indicates suboptimal behavior.

\section{Motivation and Problem Statement}
\label{motivation}

\subsection{Motivation}
\begin{figure}[!t]
    \centering
    \includegraphics[width=0.48\textwidth]{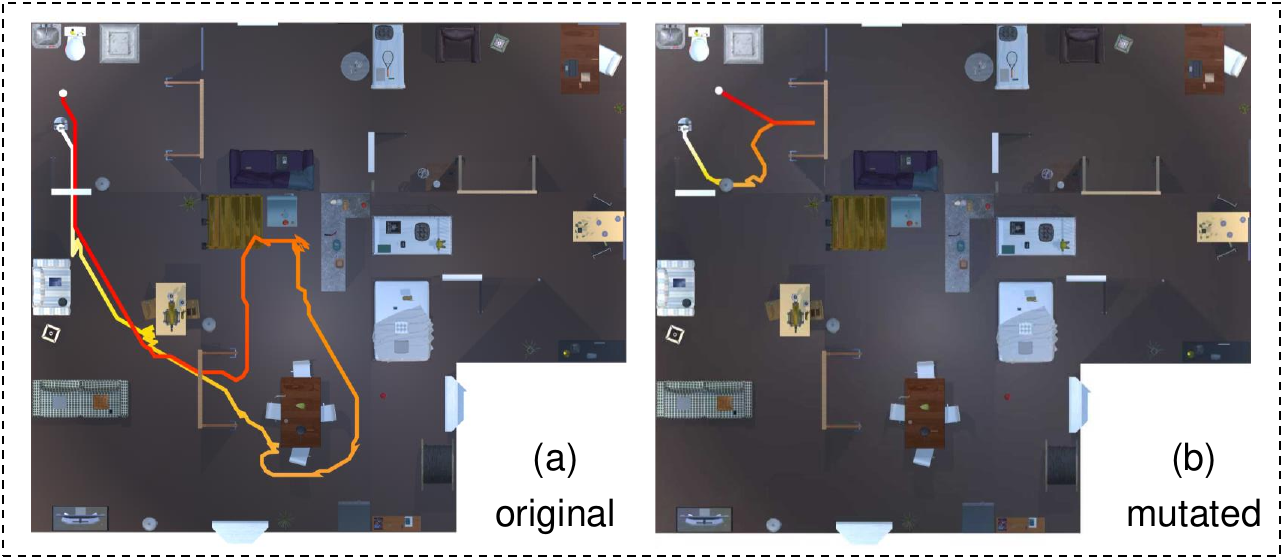} 
    \caption{A motivating example revealing an optimality issue in embodied agent decision-making. The navigation target is marked with a red circle, and the agent's terminal position is indicated by a white circle.}
    \label{fig:motivation}
\end{figure}

This section presents a real NoD case identified in SPOC~\cite{ehsani2024spoc} and demonstrates how our method can detect it.
Fig.~\ref{fig:motivation} shows the trajectories traveled by the agent under the same target and environment.
In the original scenario shown in Fig.~\ref{fig:motivation}(a), an agent is instructed to navigate to a toilet within a household environment. The agent generates a significantly circuitous path requiring 367 steps, traveling through multiple rooms with numerous redundant turns and inefficient routing decisions.
To evaluate optimality, we apply a metamorphic transformation by adding an intermediate goal.
Intuitively, this additional constraint should increase the path complexity and step count. However, as shown in Fig.~\ref{fig:motivation}(b), the agent accomplishes the modified task in only 113 steps, which is significantly fewer than the original 367 steps.
The difference between the two outcomes illustrated in Fig.~\ref{fig:motivation} reveals a clear violation: \textit{the plan generated in Fig.~\ref{fig:motivation}(a) is non-optimal since adding an intermediate goal should not result in fewer steps than the original task}. In the following sections, we introduce how to systematically generate test cases using MT to automatically discover such NoD issues.

This motivating example clearly illustrates the advantage of our approach. By defining MRs that capture the expected properties of optimal decision-making, we can automatically generate test cases to effectively evaluate an agent's planning optimality, without the need to manually define an oracle for each individual task. 


\subsection{Problem Statement}


Based on the above analysis, we formally define the research problem addressed in this paper as follows:

\noindent \textbf{Problem:}
\textit{
Given an embodied agent with task planner $\Pi$ operating in environment $E$, design a set of metamorphic relations $\mathcal{MR} = \{MR_1, MR_2, \ldots, MR_k\}$ that encode the expected behavioral properties of optimal decisions, and develop an MT-based testing framework to automatically detect NoDs in the generated action sequences through systematic test case generation guided by these MRs.}


Specifically, each metamorphic relation $MR_i$ defines the expected relationship between a source test case $(\mathcal{L}, E)$ and its corresponding follow-up test case $(\mathcal{L}^\prime, E^\prime)$. The core challenge lies in formulating MRs that capture the invariant properties of optimal planning behavior and systematically generating diverse source and follow-up test cases.

\section{Methodology}
\label{approach-details}

\subsection{Overview}

\begin{figure*}[t]
    \centering
\includegraphics[width=0.97\textwidth]{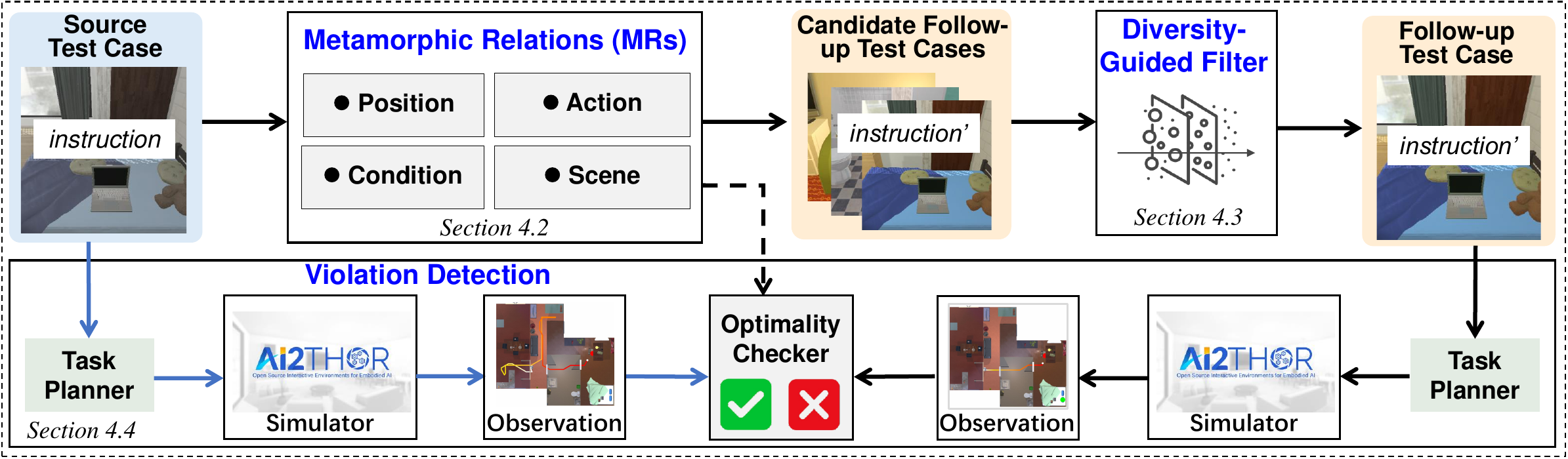} 
    \caption{The three-phase pipeline of \method: test case generation, diversity-guided selection, and violation detection.}
    \label{fig-framework}
\end{figure*}

In this section, we introduce our method \method, which operates in a three-phase pipeline as illustrated in Fig.~\ref{fig-framework}.
First, we design and formalize four types of MRs that encode the expected behavioral properties of optimal decisions. These MRs are specifically designed for action, scene, condition, and position mutations, each capturing different aspects of optimality constraints that should hold across related planning scenarios (Section~\ref{MR-def}). Given a source test case, we systematically apply these MRs to generate a set of candidate follow-up test cases.
Second, to ensure comprehensive evaluation coverage while maintaining efficiency, we propose a diversity-guided filter that selects the most valuable test cases from the candidate pool (Section~\ref{div-select}). This selection mechanism prioritizes test cases that maximize the exploration of different types of potential optimality violations, thereby enhancing the framework's ability to detect diverse categories of NoDs while avoiding redundant evaluations.
Finally, as shown in the violation detection component, \method~executes both source and the selected follow-up test cases by feeding them into the task planner and simulator to generate corresponding action sequences and execution trajectories (Section~\ref{violation-detection}). The optimality checker then performs comparisons of the execution results against the predefined MRs to identify violations, thereby detecting NoDs in the embodied agent's planning behavior.

\subsection{Metamorphic Relations (MRs)}
\label{MR-def}

\begin{figure}[!t]
    \centering
    \includegraphics[width=0.49\textwidth]{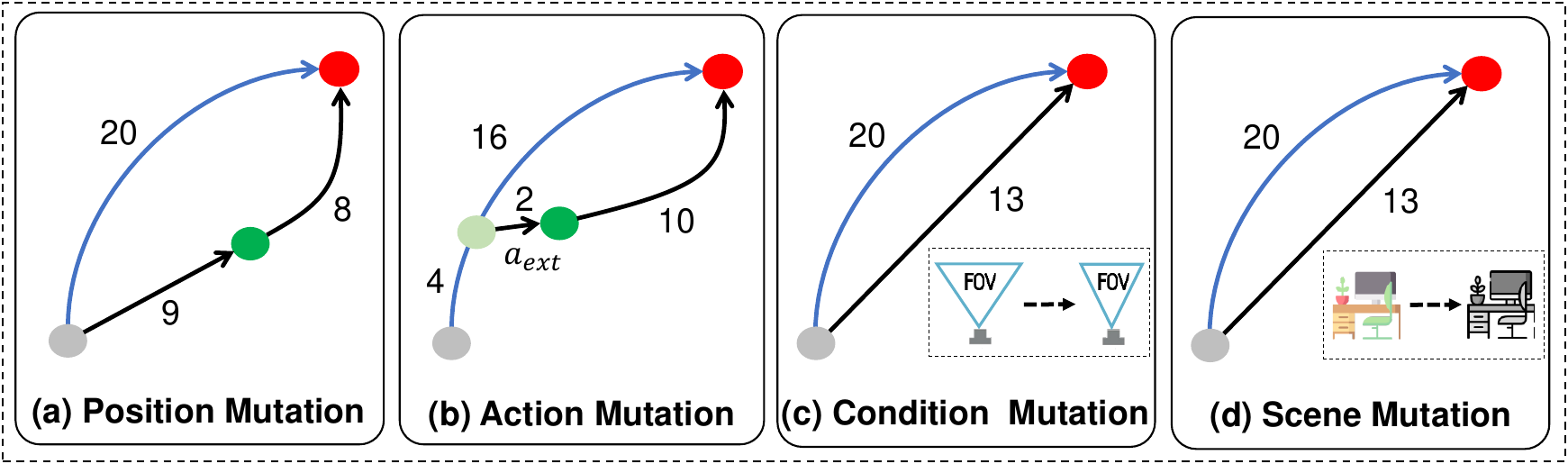} 
   \caption{Four types of mutations and their corresponding metamorphic relations for NoD detection. Gray, green, and red circles denote starting positions, intermediate waypoints, and target positions, respectively.}
    \label{fig:four-MRs}
\end{figure}

One key technical contribution of \method~is the MRs designed for NoD detection. To this end, this section designs four MRs to evaluate decision optimality in embodied agent task planning, as illustrated in Fig.~\ref{fig:four-MRs}.
The core insight behind our MR design is that optimal planners should exhibit consistent behaviors when faced with environmental or task mutations. As shown in Fig.~\ref{fig:four-MRs}, we identify four fundamental transformation categories: position, action, condition, and scene mutations, each capturing different aspects of optimality constraints that should hold across related planning scenarios.

\textbf{Position Detour Suboptimality} ($MR_{position}$).
Given an action sequence $\theta$ that successfully completes task $\mathcal{L}$ from position A to C, creating a composite path A→B→C where the agent first moves toward intermediate position B for $\eta$ steps and then continues to the original goal C should not yield a more efficient solution, i.e., $C(\theta_{A \to B}) + C(\theta_{B \to C}) \geq C(\theta_{A \to C})$.
For example, if an agent generates a plan requiring 20 steps to reach a target from position A, creating a detour path that first moves toward a different object for $\eta$ steps and then continues to the original goal should not result in a complete solution with fewer than 20 total steps.
A violation of this MR indicates that the original sequence $\theta$ was suboptimal, suggesting the planner failed to identify that an indirect initial path could lead to a more efficient overall solution to reach the target destination.


\textbf{Action Optimality Completeness} ($MR_{action}$).
Given an action sequence $\theta = \langle a_0, a_1, \ldots, a_{\tau} \rangle$ that successfully completes task $\mathcal{L}$, replacing the planned action $a_t$ at time step $t \in [1, \tau-1]$ with an extraneous action $a_{ext} \in \mathbf{A} \setminus \{a_t\}$ and re-planning the subsequent actions should not yield a more efficient solution, i.e., $C(\theta') \geq C(\theta)$ where $\theta' = \langle a_0, \ldots, a_{t-1}, a_{ext} \rangle \circ \Pi(s'_t, \mathcal{L})$.
For example, if an agent generates a navigation plan $\theta$ requiring 20 steps, replacing the action at step $t=4$ with an unrelated action like \textit{``\MakeUppercase{move arm up}"} and continuing with optimal planning from the new state should not yield a complete solution with fewer than 20 total steps.
A violation of this MR indicates that the original sequence $\theta$ lacks optimality completeness, suggesting the presence of NoDs where the planner failed to identify more efficient alternatives.

\textbf{Condition Refinement Monotonicity} ($MR_{condition}$).
Given an action sequence $\theta$ that successfully completes task $\mathcal{L}$ with completion condition $\Omega^o$, refining the completion condition to a more restrictive requirement $\Omega^r$ should not yield a more efficient solution, i.e., $C(\theta') \geq C(\theta)$ where $\theta'$ completes the refined task $\mathcal{L}'$ with condition $\Omega^r \subset \Omega^o$.
For example, if an agent generates a plan requiring 20 steps to \textit{``find the pillow"} with a Field of View (FOV) of 90 degrees, changing the condition to \textit{``pick up the pillow"} with a more restrictive FOV of 45 degrees should not result in a solution with fewer than 20 steps.
A violation of this MR indicates that the original sequence $\theta$ was suboptimal, as the planner failed to recognize that achieving the more restrictive condition could lead to a more efficient overall solution.

\textbf{Scene Invariance Property} ($MR_{scene}$).
Given an action sequence $\theta$ that successfully completes task $\mathcal{L}$ in environment $E$, modifying task-irrelevant scene elements such as object materials, lighting conditions, or decorative items should not yield a more efficient solution, i.e., $C(\theta') \geq C(\theta)$ where $\theta'$ completes the same task $\mathcal{L}$ in the modified environment $E'$ with cosmetic changes.
For example, if an agent generates a navigation plan requiring 20 steps in a room with wooden furniture and warm lighting, changing the furniture material to metal and the lighting to cool colors should not result in a solution with fewer than 20 steps.
A violation of this MR indicates that the original sequence $\theta$ was suboptimal, suggesting the planner was inappropriately influenced by task-irrelevant visual features and failed to identify the truly optimal path that should remain consistent across cosmetic scene variations.

\subsection{Diversity-Guided Filter}
\label{div-select}



Given any source test case, our four MRs can generate multiple follow-up test cases through different mutation strategies. However, applying all possible mutations can result in a large number of candidate test cases, leading to computational inefficiency and potentially redundant evaluations. The basic idea is that we want to select the most informative mutations that are likely to expose different types of NoDs while avoiding redundant test cases that explore similar aspects of optimality. To address this challenge, we introduce a diversity-guided filter component that systematically selects the most valuable test cases from the candidate pool by quantifying diversity for four types of mutations.

Specifically, for $MR_{position}$, we design a filter $F_{position}$ that measures diversity based on the perpendicular distance between the intermediate waypoint and the original trajectory. Given multiple candidate intermediate positions $\textbf{B}^{c} = \{B_1, B_2, \ldots, B_n\}$, the filter selects the waypoint that maximizes the \textit{perpendicular distance} to the original path:
\begin{equation}
B^* = \arg\max_{B_i \in \textbf{B}^{c}} Distance(B_i, \text{Path}(A \to C)).
\end{equation}
where $Distance(B_i, \text{Path}(A \to C))$ represents the shortest perpendicular distance from waypoint $B_i$ to the original direct path from A to C.

For the action mutation, we design a filter $F_{action}$ that leverages LLMs to select semantically diverse actions. Given multiple candidate alternative actions $\textbf{A}_t^{c} =\{a^1_t, a^2_t, \ldots, a^n_t\}$ and the current planned action $a_t$, we prompt the LLM to identify the action that exhibits the greatest \textit{semantic difference}.
For example, given the planned action \textit{``\MakeUppercase{move ahead}"} and candidate actions \{\textit{``\MakeUppercase{move back}"}, \textit{``\MakeUppercase{rotate right}"}, \textit{``\MakeUppercase{move arm up}"}\}, if the LLM identifies \textit{``\MakeUppercase{move back}"} as the most semantically different action, we select this action as $a_{ext}$ for the follow-up test case generation.

For the condition mutation, the filter $F_{condition}$ is defined as selecting the condition mutation with the maximum \textit{refinement divergence} from the original condition. For continuous variables, we calculate the numerical difference between conditions. For discrete variables, we leverage LLMs to identify the condition that exhibits the greatest semantic difference.
For example, when modifying the FOV from 90 degrees (original condition $\Omega^o$), we select the candidate FOV value that maximizes $|\Omega^o - \Omega^r_{c}|$, where $\Omega^r_{c}$ represents the refined condition candidates. 

Finally, for $MR_{scene}$, its filter $F_{scene}$ is implemented by calculating the \textit{scene similarity}. Among all candidates, we select the case with the lowest similarity to the original scene, ensuring that the selected follow-up test case introduces the maximum visual variation while maintaining task-irrelevance. The scene similarity is computed based on visual features extracted from the environment, including object appearances, material properties, lighting conditions, and color schemes. 
This selection strategy maximizes the diversity of visual modifications while ensuring that the changes remain semantically irrelevant to the task objective, thereby providing the maximum potential for detecting NoDs.

\subsection{Metamorphic Testing-based Violation Detection Framework}
\label{violation-detection}

This section introduces the MT-based testing framework for NoD detection. As illustrated in the bottom part of Fig.~\ref{fig-framework}, the violation detection component operates by executing both source and follow-up test cases through the task planner and simulator, then comparing their execution results against the predefined MRs.
Algorithm~\ref{alg:violation-detection} presents the complete violation detection process of \method. It takes as input a source test case and its corresponding follow-up test case selected by our diversity-guided filter, the task planner under test, and the specific MR to be evaluated. The algorithm outputs a violation detection result indicating whether a NoD has been identified.

\begin{algorithm}[t]
\caption{Metamorphic Testing-based Violation Detection}
\label{alg:violation-detection}
\begin{algorithmic}[0]  %
\STATE \textbf{Input:} Source case $(\mathcal{L}, E)$, follow-up case $(\mathcal{L}', E')$, planner $\Pi$, and a specific metamorphic relation $MR_k \in \mathcal{MR}$.
\STATE \textbf{Output:} Violation detection result.
\end{algorithmic}
\begin{algorithmic}[1]
\STATE Initialize $violated \leftarrow False$; \label{init-violation}
\STATE Execute source case: $\theta_{s} \leftarrow \Pi(\mathcal{L}, E)$; \label{exe-source}
\STATE Execute follow-up case: $\theta_{f} \leftarrow \Pi(\mathcal{L}', E')$; \label{exe-follow}
\IF{either task fails} \label{failure-start}
    \STATE \textbf{return} \texttt{Task\_Execution\_Failed};
\ENDIF \label{failure-end}
\STATE Calculate costs: $C_{s} \leftarrow C(\theta_{s})$, $C_{f} \leftarrow C(\theta_{f})$; \label{cost-calculate}
\IF{$C_{f} < C_{s}$} \label{checker-start}
    \STATE $violated \leftarrow True$;\label{presence-nod}
\ENDIF \label{checker-end}
\STATE \textbf{return} $\langle violated, MR_k, C_{s}, C_{f} \rangle$;
\end{algorithmic}
\end{algorithm}

The algorithm begins with initialization (Line~\ref{init-violation}), setting the violation flag to $False$ to track whether a NoD violation is detected during the execution process.
Lines~\ref{exe-source}--\ref{exe-follow} describe the task execution phase. \method~begins by executing both the source and follow-up test cases using the task planner~$\Pi$ to generate their respective action sequences $\theta_s$ and $\theta_f$ in the simulator. If either task execution fails, the algorithm immediately returns a task execution failure status (Lines~\ref{failure-start}--\ref{failure-end}), as meaningful optimality comparison cannot be performed without successful task completion.
Following successful execution, the algorithm calculates the cost metrics for both action sequences using the cost function $C(\cdot)$ (Line~\ref{cost-calculate}). Lines~\ref{checker-start}--\ref{checker-end} implement the optimality checker, which examines whether the transformed scenario yields superior efficiency ($C_f < C_s$). When a violation is detected, the violated flag is set to $True$ (Line~\ref{presence-nod}), indicating the presence of a NoD.
Finally, the algorithm returns a comprehensive result tuple containing the violation status, the specific MR tested, and both cost values.
\section{Evaluation}
\label{experiment-sec}

In this section, we evaluate the effectiveness and efficiency of \method~in NoD detection.
We aim to answer the following Research Questions (RQs):
\begin{itemize}
\item \textbf{RQ1:} How effective is the proposed \method~in detecting NoDs?
\item \textbf{RQ2:} How useful is the proposed diversity-guided filter in enhancing the efficiency of \method?
\item \textbf{RQ3:} How does \method~perform in comparison with other existing testing methods in terms of NoD detection capability?
\end{itemize}
For RQ1, we evaluate \method~in detecting NoDs across various task categories with four well-trained task planning models in the AI2-THOR simulator (Section~\ref{sec:rq1}).
To answer RQ2, we perform ablation studies to compare NoD detection performance and diversity coverage across the four MRs with and without the diversity-guided filter (Section~\ref{sec:rq2}).
To answer RQ3, we compare \method~with four baseline methods (Section~\ref{sec:rq3}).

\begin{table*}
\caption{Violation Detection Rates of \method Across Different Task Complexity and Planning Models.}
\label{tab:effectiveness}
\tabcolsep 5.2pt
\centering
\begin{tabular}{c|c|ccc|ccc|ccc|ccc}  
\hline 
\multirow{2}{*}{\textbf{Type}} & \multirow{2}{*}{\textbf{Model}} & \multicolumn{3}{c|}{\textbf{Position}} & \multicolumn{3}{c|}{\textbf{Action}} & \multicolumn{3}{c|}{\textbf{Condition}} & \multicolumn{3}{c}{\textbf{Scene}} \\ \cline{3-14}
 &  & \textbf{Long} & \textbf{Medium} & \textbf{Short} & \textbf{Long} & \textbf{Medium} & \textbf{Short} & \textbf{Long} & \textbf{Medium} & \textbf{Short} & \textbf{Long} & \textbf{Medium} & \textbf{Short} \\ \hline
\multirow{4}{*}{Slight} & DINOv2& 0.0\% & 0.0\% & 5.3\% & 5.9\% & 0.0\% & 10.4\% & 6.3\% & 8.3\% & 4.7\% & 22.2\% & 11.4\% & 6.8\% \\
 & SigLIP& 7.1\% & 2.6\% & 6.8\% & \textbf{19.2\%} & 5.3\% & 9.8\% & 3.8\% & 2.7\% & 14.3\% & \textbf{29.2\%} & 5.6\% & 12.2\% \\
 & SigLIP-Nav & 0.0\% & 5.4\% & 2.3\% & 13.6\% & 5.4\% & 10.9\% & 8.0\% & 5.3\% & 4.4\% & 5.0\% & 26.5\% & 10.9\% \\
 & SigLIP-Det & 15.6\% & 10.2\% & 9.8\% & 14.3\% & \textbf{15.3\%} & \textbf{12.9\%} & 4.2\% & 10.3\% & 14.5\% & 12.5\% & 13.8\% & 9.7\% \\ \hline
\multirow{4}{*}{Moderate} & DINOv2& 0.0\% & 9.4\% & 7.9\% & 0.0\% & 8.1\% & 8.3\% & 18.8\% & 2.8\% & 2.3\% & 5.6\% & 2.9\% & 2.3\% \\
 & SigLIP& 0.0\% & 10.3\% & 4.5\% & 7.7\% & 5.3\% & 2.4\% & 0.0\% & 8.1\% & 11.9\% & 4.2\% & 2.8\% & 14.6\% \\
 & SigLIP-Nav & 4.5\% & 2.7\% & 2.3\% & 9.1\% & 10.8\% & 2.2\% & 4.0\% & 10.5\% & 4.4\% & 5.0\% & 2.9\% & 6.5\% \\
 & SigLIP-Det & 0.0\% & 13.6\% & 11.5\% & 2.4\% & 1.7\% & 6.5\% & 4.2\% & 8.6\% & 16.1\% & 8.3\% & 6.9\% & 4.8\% \\ \hline
\multirow{4}{*}{Severe} & DINOv2& 8.7\% & 15.6\% & 18.4\% & 5.9\% & 5.4\% & 0.0\% & 18.8\% & 16.7\% & 25.6\% & 5.6\% & 22.9\% & \textbf{22.7\%} \\
 & SigLIP& 7.1\% & 12.8\% & 25.0\% & 7.7\% & 5.3\% & 2.4\% & 19.2\% & \textbf{32.4\%} & 14.3\% & 8.3\% & \textbf{27.8\%} & 17.1\% \\
 & SigLIP-Nav & 13.6\% & \textbf{29.7\%} & 32.6\% & 4.5\% & 2.7\% & 6.5\% & \textbf{28.0\%} & 28.9\% & \textbf{28.9\%} & 10.0\% & 14.7\% & 21.7\% \\
 & SigLIP-Det & \textbf{17.8}\% & 22.0\% & \textbf{45.9\%} & 7.1\% & 8.5\% & 4.8\% & 14.6\% & 15.5\% & 19.4\% & 18.8\% & 20.7\% & 14.5\% \\ \hline
\multirow{4}{*}{Total} & DINOv2& 8.7\% & 25.0\% & 31.6\% & 11.8\% & 13.5\% & 18.7\% & 43.8\% & 27.8\% & 32.6\% & 33.3\% & 37.1\% & 31.8\% \\
 & SigLIP& 14.3\% & 25.6\% & 36.4\% & 34.6\% & 15.8\% & 14.6\% & 23.1\% & 43.2\% & 40.5\% & 41.7\% & 36.1\% & 43.9\% \\
 & SigLIP-Nav & 18.2\% & 37.8\% & 37.2\% & 27.3\% & 18.9\% & 19.6\% & 40.0\% & 44.7\% & 37.8\% & 20.0\% & 44.1\% & 39.1\% \\
 & SigLIP-Det & 33.3\% & 45.8\% & 67.2\% & 23.8\% & 25.4\% & 24.2\% & 22.9\% & 34.5\% & 50.0\% & 39.6\% & 41.4\% & 29.0\% \\ \hline
\end{tabular}
\end{table*}

\subsection{Experimental Settings}

In the following, we provide a detailed description of the experimental setup, including the model configurations, implementation details, and evaluation metrics.




\textbf{Model Selection.} 
Our experiments employ four state-of-the-art pre-trained models from~\cite{ehsani2024spoc} as the task planner: DINOv2, SigLIP, SigLIP-Nav, and SigLIP-Det. 
Specifically, DINOv2, SigLIP, and SigLIP-Det are trained on the \bench{}\fifteen  dataset~\cite{ehsani2024spoc}, while SigLIP-Nav is trained on the \benchnav{}\fifteen dataset~\cite{ehsani2024spoc}. SigLIP, SigLIP-Nav, and SigLIP-Det utilize \siglip~\cite{zhai2023sigmoid} image and text encoders, while DINOv2 employs \dino~\cite{oquab2023dinov2} image and text encoders. SigLIP-Det incorporates ground truth detection capabilities provided by the simulator. These models represent different approaches to visual-language understanding and navigation planning, enabling comprehensive evaluation of our NoD detection framework across diverse planning paradigms.

\textbf{Implementation details.} We implement four MRs in the AI2-THOR simulator. For $MR_{position}$, we assign a different target object for the planner in the first several steps (1/4 of the entire step in the source test case) and recovering the original goal in the following steps. For $MR_{action}$, we utilize GPT-4~\cite{achiam2023gpt} to determine the semantic distance and substitute the action at the middle point of the original action sequence, i.e., $t = \tau/2$ .
For $MR_{condition}$, we adjust the FOV of the agent's camera with scaling factors ranging from 0.75 to 1.0. 
For $MR_{scene}$, we generate different scenes through two approaches: (1) randomizing scene materials with sensible alternatives from similar material types, and (2) randomizing lighting parameters including color, hue (0 $\sim$ 1), and brightness (0.5 $\sim$ 1.5).
We follow the official evaluation split of the \bench{}\fifteen \ObjectNav benchmark~\cite{ehsani2024spoc} to generate the source test case set. We categorize the test set into three task complexity levels: \textit{Short}, \textit{Medium}, and \textit{Long}, based on the step count of feasible reference paths provided in the dataset.



\textbf{Evaluation Metrics.} 
We measure the effectiveness of MRs through violation detection rate, which represents the percentage of test cases that result in detected violations by \method~out of the total number of executed test cases.
In addition, we define a metric, Violation Severity (\textit{VS}), to quantify the degree of suboptimality detected in each violation case: $\textit{VS} = (L_{orig}-L_{mut})/L_{orig}$, where $L_{orig}$ and $L_{mut}$ denote the number of steps in the agent plans of the source case and the mutation case, respectively.
We classify the violation cases into three levels: \textit{Severe} ($VS>20\%$), \textit{Moderate} ($10\% < VS \le 20\%$), and \textit{Slight}($VS\le 10\%$) based on the step count reduction between the mutation case and the original case.

\subsection{RQ1: Effectiveness of \method}\label{sec:rq1}

\subsubsection{Detection Performance Evaluation}

To evaluate the effectiveness of \method,
we applied each MR to 400 test cases (200 source test cases and 200 follow-up test cases) across four planners.
Table~\ref{tab:effectiveness} shows the violation detection rates of our four MRs across task complexity categories.
Table~\ref{tab:effectiveness} presents the violation detection rates across different VS levels (Slight, Moderate, Severe, and Total), MR types (Position, Action, Condition, Scene), and test case lengths (Long, Medium, Short).

From Table \ref{tab:effectiveness}, it can be observed that across all cases, despite variations in detection rates among different mutation types, they all maintain a ``rising with severity" trend: the detection rate of \textit{Severe} violations is consistently the highest, followed by \textit{Moderate} violations, and \textit{Slight} violations have the lowest rates. For example, in Position mutation (Short-term case), SigLIP-Det achieves a 45.9\% detection rate for \textit{Severe} violations, far exceeding its 9.8\% rate for \textit{Slight} violations. This indicates that our proposed four MRs can effectively identify NoDs of varying severity, with a particular ability to detect severe optimality violations.

Among the four MRs, $MR_{\text{condition}}$ and $MR_{\text{position}}$ demonstrate prominent performance. For example, $MR_{\text{condition}}$ of SigLIP-Nav achieves detection rates of 28.0\% (\textit{Severe}, Long-term case), 10.5\% (\textit{Moderate}, Medium-term case), and 8\% (\textit{Slight}, Long-term case), while $MR_{\text{position}}$ excels in detecting \textit{Severe} violations (e.g., 45.9\% for SigLIP-Det in Short-term cases). In contrast, $MR_{\text{Action}}$ exhibits relatively lower but still meaningful detection rates (e.g., 19.2\% for SigLIP in \textit{Slight} violations with Long-term cases). These disparities arise from the distinct mechanisms of each MR: $MR_{\text{condition}}$ tightens task completion requirements (e.g., restricting FOV to 75\%), and if fewer steps are used after mutation, it violates the monotonicity constraint (stricter conditions should not enable more efficient solutions), effectively capturing such violations (e.g., 28.9\% for SigLIP-Nav in Severe violations with Short-term cases); $MR_{\text{scene}}$ selects the least similar scene for mutation, targeting behaviors that rely on irrelevant details, which violate its invariance constraint; $MR_{\text{position}}$ selects a new target maximally different from the original, and if the agent completes the task in fewer steps after the detour, it violates $C(\theta_{A \to B}) + C(\theta_{B \to C}) \geq C(\theta_{A \to C})$, with the highest total detection rate among models reaching 67.2\%; $MR_{\text{Action}}$ amplifies planners’ vulnerabilities in action understanding and strategy adaptability by employing semantically conflicting actions and each of the four models has a total violation detection rate of over 20\%. Thus, each MR targets distinct aspects of optimality, their combined performance attests to the method’s comprehensiveness, and the lack of a single dominant MR shows they complement one another to cover diverse NoDs.

All four models exhibit the same severity-dependent trend. SigLIP-Det generally achieves the highest detection rates (e.g., 67.2\% total detection rate for $MR_{position}$ in Short-term cases), while DINOv2 shows more moderate but still significant rates (e.g., 31.8\% total detection rate for $MR_{scene}$ in Short-term cases). This demonstrates that our MRs method is robustly effective across different models, confirming its generality as a NoD detection framework. MRs target fundamental optimality properties of embodied task planning (invariance across positions, actions, conditions, and scenes) rather than model-specific behaviors. Even models with strong generalization capabilities (such as DINOv2) cannot avoid violations when undergo NoD detection, ensuring the MRs remain effective. The higher detection rate of SigLIP-Det may stem from its focus on object detection and path localization, making it more sensitive to $MR_{position}$ and $MR_{scene}$ further verifying that MRs can exploit model-specific vulnerabilities while maintaining broad applicability.

In summary, the experimental results confirm the effectiveness of the four MRs: they can stably detect NoDs across different \textit{VS} categories, task lengths, and models, with performance tightly aligned to their design logic of targeting core optimality invariants via diversity-guided mutations.

\subsubsection{Violation Pattern Analysis}

\begin{figure*}[!t]
    \centering
\includegraphics[width=0.93\textwidth]{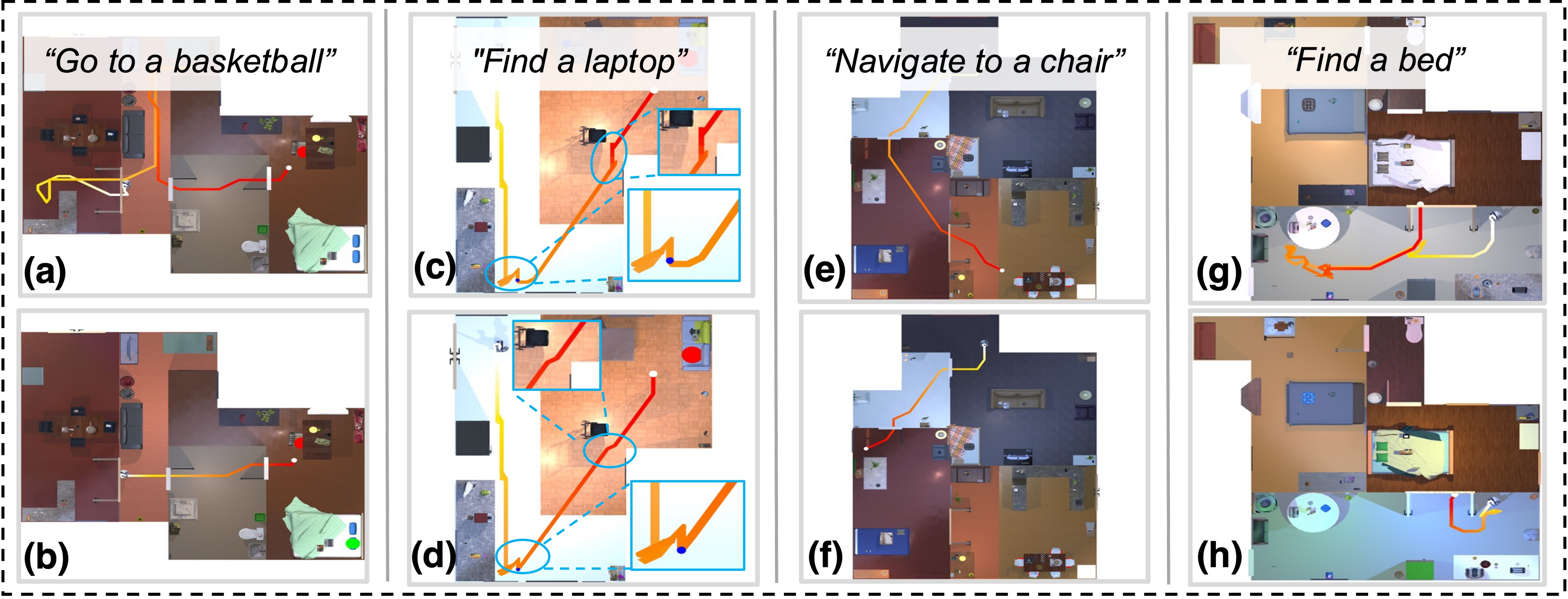} 
    \caption{Typical failure cases of NoD detected by \method.}
    \label{fig-case}
\end{figure*}

To gain a systematic and comprehensive understanding of the NoDs revealed by \method, we inspect MR violations and grouped them into four categories for in-depth exploration. Fig. \ref{fig-case} reports examples of issues detected by \method. These four examples are respectively refer to $MR_{position}$, $MR_{action}$, $MR_{conditon}$, and $MR_{scene}$ from left to right.

\textbf{Inefficient Global Exploration Strategy.} Fig. \ref{fig-case}(a) shows the source path, and Fig. \ref{fig-case}(b) shows the path with $MR_{position}$, applied. In Fig. \ref{fig-case}(b), the agent navigates toward the farthest target from the original goal (``go to a pillow") for the first 1/4 of the path. This enables broader early exploration and more comprehensive environmental information acquisition. Upon switching back to the original goal at the 1/4 mark, the agent, leveraging prior global environmental awareness, plans a more direct, obstacle-avoiding path—thus skipping an extra room during exploration and shortening the route. In contrast, Fig. \ref{fig-case} (a), constrained by limited information, exhibits a detoured path.

\textbf{Local Optimal Action Sequence Trapping.} 
The second pattern involves agents becoming trapped in locally optimal action sequences. 
Fig.~\ref{fig-case}(c) shows the source path, and Fig.~\ref{fig-case}(d) shows the path with $MR_{action}$ applied. In Fig.~\ref{fig-case}(d), we replace the original left-turn action with a right-turn. As shown, Fig.~\ref{fig-case}(d) eliminates a detour in the middle section compared to Fig.~\ref{fig-case}(c), resulting in a shorter route (over 40 steps fewer than the original result).  In Fig.~\ref{fig-case}(c), the agent encountered a corner at the room edge, which acted as a distraction and led to a longer path. In Fig.~\ref{fig-case}(d), the reversed action altered the traveling direction, avoiding this distraction, and the replanned path led more directly to the target.

\textbf{Inadequate Environmental Perception.} Fig.~\ref{fig-case}(e) shows the source path, Fig.~\ref{fig-case}(f) is the path with $MR_{conditon}$ applied, which the agent's FOV reduced to 75\% of the source. In Fig.~\ref{fig-case}(f), the agent detects the chair at a closer distance, whereas in Fig.~\ref{fig-case}(e), it fails to detect the chair initially and continues exploring until a farther location. This is because with 75\% FOV, the agent uses a more aggressive search strategy, prioritizing rapid exploration toward the target's general direction. Meanwhile, the smaller focus area increases the target chair's proportion in the sensor's FOV and enhances its contrast with the surroundings, enabling easier recognition, quicker positioning, and shorter path planning. In contrast, the algorithm of
Fig.~\ref{fig-case}(e) tends to adopt a conservative global exploration strategy, requiring full environmental understanding before path planning—lengthening the exploration process.

\textbf{Poor Environmental Adaptation.} Fig.~\ref{fig-case}(g) shows the source path, while Fig.~\ref{fig-case}(h) presents the path after using $MR_{scene}$. It can be seen that the path length of Fig.~\ref{fig-case}(h) (44 steps) is much shorter than that of Fig.~\ref{fig-case}(g) (149 steps). Although the light in Fig.~\ref{fig-case}(h) is dimmer, the contrast between the bed and the surrounding environment is enhanced, making it easier for the agent to identify and locate the bed. In Fig.~\ref{fig-case}(g), the light is relatively strong, resulting in low distinguishability between the bed and the background, and the sensor imaging is blurred. Therefore, when the agent first enters the room with the bed, it fails to capture the bed, and the agent needs to spend more time searching and confirming the target position, which leads to a longer route. 

The four MRs all reveal non-optimal decisions in embodied agents by introducing intentional changes to the original scenario, such as altering initial positions, reversing actions, adjusting sensory conditions, or modifying environmental settings. In each case, the modified scenario (via MR) results in a shorter, more efficient path than the original, exposing the agent’s suboptimal planning in the source case. These inconsistencies arise because the agent fails to adapt effectively to environmental constraints, sensory limitations, or action choices, highlighting flaws in its decision-making logic when navigating toward targets.


\subsection{RQ2: Usefulness of Selection Strategy}\label{sec:rq2}
To measure the diversity of violation cases, we apply spectral clustering~\cite{ng2001spectral} using Euler distance for initial views and Levenshtein distance for action sequences, with optimal cluster numbers determined by maximizing the Silhouette Coefficient~\cite{shahapure2020cluster}. 
 The overall violation diversity (\textit{VD}) is then defined as the total number of unique cluster pairs \((c_{\text{view}}, c_{\text{action}})\) identified across all cases, where \(c_{\text{view}}\) and \(c_{\text{action}}\) denote the cluster assignments based on the initial view and action sequence, respectively:
\begin{equation}
\textit{VD} = \left| \left\{ (c_{\text{view}}^i, c_{\text{action}}^i) \mid i = 1, 2, \ldots, N \right\} \right|
\end{equation}
where \(N\) is the total number of violation cases, and \(|\cdot|\) denotes the cardinality of the set of unique cluster pairs.

Table~\ref{tab:table4} compares violation detection rates and violation diversity scores between filtered and unfiltered approaches across different models and MRs.
For $MR_{\text{position}}$, the filter achieves enhancement levels of 4.3\%, 0.9\%, 6.9\%, and 6.7\% for DINOv2, SigLIP, SigLIP-Nav, and SigLIP-Det, respectively. SigLIP exhibits minimal enhancement (0.9\%), while SigLIP-Nav and SigLIP-Det achieve substantial improvements (6.9\% and 6.7\%). This discrepancy can be attributed to two factors: (1) SigLIP-Nav's extended training on navigation tasks enhances object recognition accuracy, and (2) SigLIP-Det's ground truth detector increases sensitivity to target position changes, enabling more precise mutation execution and eliminating noise from inaccurate target localization. Consequently, position mutation achieves both higher detection rates on these models (50.3\% for SigLIP-Det, 33.3\% for SigLIP-Nav) and stronger filter effectiveness.
Interestingly, SigLIP, which shows minimal violation rate enhancement, exhibits the highest diversity improvement (increasing from 20 to 24 clusters). This suggests that while weak object localization may introduce noise during mutation, it contributes to generating novel trajectories. The inverse relationship between diversity and violation rate improvements, despite consistent underlying patterns, indicates that models with stronger localization capabilities discover violations with both higher quantity and quality.

\begin{table*}[!t]
\caption{Violation Detection Rate Comparison of \method~with and without the Diversity-Guided Filter}
\label{tab:table4}
\tabcolsep 2.2pt
\footnotesize
\centering
\begin{tabular}{c|c|ccc|ccc|ccc|ccc|ccc}
\hline
\multirow{2}{*}{\textbf{Metric}} & \multirow{2}{*}{\textbf{Model}} & \multicolumn{3}{c|}{\textbf{Position}} & \multicolumn{3}{c|}{\textbf{Action}} & \multicolumn{3}{c|}{\textbf{Condition}} & \multicolumn{3}{c|}{\textbf{Scene}} & \multicolumn{3}{c}{\textbf{Average}} \\ \cline{3-17}
 & & \textbf{Filter} & \textbf{w/o} & \textbf{Enhance} & \textbf{Filter} & \textbf{w/o} & \textbf{Enhance} & \textbf{Filter} & \textbf{w/o} & \textbf{Enhance} & \textbf{Filter} & \textbf{w/o} & \textbf{Enhance} & \textbf{Filter} & \textbf{w/o} & \textbf{Enhance}\\ \hline
\multirow{4}{*}{Violation} & DINOv2& 23.7\% & 19.4\% & 4.3\% & 15.7\% & 13.7\% & 2.0\% & 32.6\% & 28.4\% & 4.2\% & 34.0\% & 25.8\% & \textbf{8.2\%} & 26.5\% & 21.8\%	& 4.7\% \\
 & SigLIP& 27.0\% & 26.1\% & 0.9\% & 20.0\% & 19.0\% & 1.0\% & 37.1\% & 29.5\% & \textbf{7.6\%} & \textbf{40.6\%} & 35.6\% & 5.0\% & 31.2\% &	27.6\%	& 3.6\% \\
 & SigLIP-Nav & 33.3\% & 26.5\% & \textbf{6.9\%} & 21.0\% & 18.1\% & 2.9\% & \textbf{40.7\%} & 36.1\% & 4.6\% & 37.0\% & 35.0\% & 2.0\% &33.0\%	& 28.9\%	& 4.1\% \\
 & SigLIP-Det & \textbf{50.3\%} & 43.6\% & 6.7\% & \textbf{24.5\%} & 19.0\% & \textbf{5.5\%} & 36.9\% & 32.1\% & 4.8\% & 36.3\% & 33.9\% & 2.4\% & \textbf{37.0\%} & 32.2\%	& \textbf{4.8\%} \\ \hline
\multirow{4}{*}{Diversity} & DINOv2& 16 & 14 & 2 & 14 & 10 & \textbf{4} & 18 & 16 & 2 & 18 & 16 & 2 & 16.5	& 14 & 2.5 \\
 & SigLIP& 24 & 20 & \textbf{4} & 14 & 13 & 1 & 24 & 18 & \textbf{6} & 21 & 18 & 3 & 20.75 & 17.25 & 3.5\\
 & SigLIP-Nav & 25 & 22 & 3 & 16 & 12 & \textbf{4} & 24 & 18 & \textbf{6} & \textbf{23} & 18 & \textbf{5} & 22 & 17.5 & \textbf{4.5}\\
 & SigLIP-Det & \textbf{35} & 33 & 2 & \textbf{19} & 15 & \textbf{4} & \textbf{30} & 25 & 5 & 22 & 22 & 0 & \textbf{26.5} & 23.75 & 2.75\\ \hline
\end{tabular}
\end{table*}

Compared to other mutations, $MR_{action}$ exhibits the shortest duration and most conservative mutation characteristics, with detection rates of 15.7\%, 20.0\%, 21.0\%, and 24.2\% across the four models. The filter mechanism demonstrates more pronounced detection rate enhancement on the SigLIP-Det model (5.5\%) compared to other models, indicating that extraneous action perturbations have more significant impacts on subsequent behaviors of higher-performing models. Models with long-range planning capabilities can make more reasonable judgments about perturbed situations, leading to more efficient path planning. Therefore, applying stronger perturbations (filter mechanism) facilitates the discovery of model non-optimality. Similarly, since this assistance manifests in subsequent long-range planning, semantically distant actions generate more diverse subsequent estimations compared to random actions, resulting in action mutation's diversity improvement trend aligning with violation rate enhancement trends.

For $MR_{condition}$, the filter mechanism achieves the highest detection rate improvement on SigLIP (37.1\%, 7.6\% enhancement) and the lowest on DINOv2 (32.6\%, 4.2\% enhancement), indicating that different image encoders exhibit distinct characteristics under receptive field perturbations. Similarly, filter mechanism diversity improvements are highest on SigLIP (6 clusters) and DINOv2 the lowest (2 clusters). SigLIP's image encoder outperforms DINOv2 on the \bench{}\fifteen test set, suggesting that higher-performing models exhibit greater sensitivity to receptive field reduction perturbations. Consequently, after narrowing the field of view, models have increased opportunities to discover objects that the original path failed to focus on, leading to more direct paths and potentially missed targets. Balancing model performance and violation discovery efficiency remains an important research question.

$MR_{scene}$ primarily targets model environmental generalization perception through perturbations. Violation detection rates on SigLIP (40.6\%) exceed those on DINOv2 (34.0\%), indicating that despite DINOv2's inferior performance compared to SigLIP on specific task, it exhibits stronger generalization capabilities and more consistent actions across different scenarios for the same task. 
However, we note minimal filter effect on the SigLIP-Det model for $MR_{scene}$ of violation diversity, which can be attributed to the model's tolerance to scene variations. 



The diversity-guided filter demonstrates consistent effectiveness across all models and MRs. As shown in Table~\ref{tab:table4}, the filter enhances violation detection rates by an average of 4.7\%, 3.6\%, 4.1\%, and 4.8\% for DINOv2, SigLIP, SigLIP-Nav, and SigLIP-Det, respectively, while simultaneously improving diversity scores by 2.5, 3.5, 4.5, and 2.75 clusters. This dual enhancement is particularly significant because it addresses both quantity (detection efficiency) and quality (coverage breadth) dimensions of NoD detection.
The consistent improvements across different model architectures and capabilities validate the generalizability of our selection strategy. 
The varying enhancement patterns across MRs provide insights into the relationship between model characteristics and mutation effectiveness. Models with stronger perception capabilities (e.g., accurate localization, better image encoders) show greater improvements under position and condition mutations, while models with advanced planning capabilities benefit more from action mutations. These findings demonstrate that our diversity-guided approach not only improves detection comprehensiveness but also reveals model-specific strengths and weaknesses.


\subsection{RQ3: Comparison with Other Testing Methods} \label{sec:rq3}
To demonstrate the advantages of \method, we compare it with six baseline methods: two Random Testing (RT) methods, two Property-Based Testing (PBT) methods, and two traditional MT methods. 
The two RT methods generate perturbed model weights to search for shorter paths: RT-GF employs Gaussian Fuzzing, while RT-NAI utilizes Neuron Activation Inverse~\cite{ma2018deepmutation}. For each test case, if the perturbed model produces a shorter path than the original model, a NoD is identified.
The two PBT methods exploit specific properties of optimal planning. PBT-NR (No Revisit) leverages the principle that optimal planners should not revisit the same spatial grid more than twice. PBT-SP (Sub-Problems) constructs sub-problems by initializing the agent at intermediate positions along the original trajectory, exploiting the property that sub-problem path lengths should not exceed the corresponding segments in the original full-problem solution.
We adapt two traditional MT approaches to embodied navigation. MT-Obstacle, adapted from MRIP2~\cite{ayerdi2022performance}, introduces additional obstacles to the environment to test planning robustness. MT-Transform, adapted from~\cite{zhang2019testing}, applies three types of transformations (Points-related, Threat-environments-related, and Geometric Transformation) to evaluate planning consistency across environmental variations.

Table \ref{tab:table6} presents a comparative analysis of the NoD detection performance between our proposed MRs-based method (encompassing four MRs: position, action, condition, and scene) and baseline testing approaches.
The results demonstrate the superiority of our MRs method. Across all models, our method achieves competitive detection rates with average performances of 33.6\% (Position), 20.3\% (Action), 36.9\% (Condition), and 37.0\% (Scene), respectively. In contrast, the six baseline methods exhibit significantly lower detection rates: the average rates for MT-Obstacle, MT-Transform, PBT-SP, PBT-NR, RT-GF, and RT-NAI are 20.3\%, 27.3\%, 12.2\%, 5.1\%, 2.2\%, and 3.2\%, respectively. Overall, our four well-designed MRs achieve an average violation detection rate of 31.9\%, which is substantially higher than that of the baseline methods.

\begin{table*}
\caption{NoD Violation Detection Performance Comparison with Different Baseline Methods}
\label{tab:table6}
\centering
\begin{tabular}{c|cccc|cc|cc|cc}
\hline
\multirow{2}{*}{\textbf{Testing Method}} & \multicolumn{4}{c|}{\multirow{2}{*}{\textbf{\method}}} & \multicolumn{6}{c}{\textbf{Baseline Methods}}\\
\cline{6-11}
& & & & &  \multicolumn{2}{c|}{\textbf{Metamorphic Testing}} & \multicolumn{2}{c|}{\textbf{Property-Based Testing}} & \multicolumn{2}{c}{\textbf{Random Testing}} \\ \hline
\textbf{Model} & \textbf{Position} & \textbf{Action} & \textbf{Condition} & \textbf{Scene} & \textbf{MT-Obstacle} & \textbf{MT-Transform} & \textbf{PBT-SP} & \textbf{PBT-NR} & \textbf{RT-GF} & \textbf{RT-NAI} \\ \hline
DINOv2 & 23.7\% & 15.7\% & 32.6\% & 34.0\% & 14.1\% & 19.8\% & 8.6\% & 3.33\% & 1.1\% & 0.0\% \\
SigLIP & 27.0\% & 20.0\% & 37.1\% & \textbf{40.6\%} & 17.5\% & 26.4\% & 12.0\% & 3.81\% & 0.0\% & 0.0\% \\
SigLIP-Nav & 33.3\% & 21.0\% & \textbf{40.7\%} & 37.0\% & \textbf{25.2\%} & 28.9\% & \textbf{14.3\%} & 6.67\% & 0.0\% & 0.0\% \\
SigLIP-Det & \textbf{50.3\%} & \textbf{24.5\%} & 36.9\% & 36.3\% & 24.3\% & \textbf{33.9\%} & 13.8\% & \textbf{6.75\%} & \textbf{7.8\%} & \textbf{12.7\%} \\ \hline
\multirow{2}{*}{Average} & 33.6\% & 20.3\% & 36.9\% & 37.0\% & 
\multirow{2}{*}{20.3\%} & \multirow{2}{*}{27.3\%} & \multirow{2}{*}{12.2\%} & \multirow{2}{*}{5.1\%} & \multirow{2}{*}{2.2\%} & \multirow{2}{*}{3.2\%}
\\ \cline{2-5} 
 & \multicolumn{4}{c|}{\textbf{31.9\%}} &  &  &  &  &  &  \\ \hline
\end{tabular}
\end{table*}

Among our four MRs, Scene and Condition achieve the highest detection rates (37.0\% and 36.9\%), indicating that environmental and perceptual variations are particularly effective at exposing planning inefficiencies. Position MR shows strong performance (33.6\%), especially on SigLIP-Det (50.3\%), which benefits from ground truth detection that enhances sensitivity to spatial relationships. Action MR exhibits moderate performance (20.3\%), as action-level perturbations require models with sophisticated long-horizon planning capabilities to reveal non-optimality.

Across the four models evaluated, SigLIP-Det achieves the highest overall detection rates, particularly for Position MR (50.3\%). Its ground truth detection capability makes it more sensitive to spatial perturbations, facilitating NoD detection. SigLIP-Nav shows strong performance on Condition MR (40.7\%), benefiting from its extended training on navigation tasks. SigLIP demonstrates high detection rates for Scene MR (40.6\%), indicating sensitivity to environmental variations. DINOv2 exhibits relatively lower detection rates across all MRs, likely due to stronger generalization capabilities that maintain robust performance despite environmental perturbations.

For the baseline methods, MT-Transform (27.3\%) performs best, as its transformation-based approach shares conceptual similarities with our MR design. However, it lacks the systematic diversity-guided selection, resulting in lower effectiveness. MT-Obstacle (20.3\%) shows moderate performance but is limited to obstacle-related scenarios. PBT-SP (12.2\%) outperforms PBT-NR (5.1\%) because sub-problem decomposition aligns better with hierarchical planning, particularly benefiting SigLIP-Nav (14.3\%). PBT-NR's single rule proves insufficient for capturing diverse NoD patterns. RT methods perform poorly overall (RT-GF: 2.2\%, RT-NAI: 3.2\%), as weight perturbations lack systematic targeting of planning inefficiencies. 
The comparison demonstrates the superiority of our proposed method over baseline approaches. Our method consistently outperforms all baselines across different model architectures and MR types, showing substantial advantages over property-based and random testing methods. These results validate that systematic metamorphic testing is substantially more effective for NoD detection than alternative strategies.

\subsection{Threats to Validity}




Limitations in experimental design and environmental factors may affect validity. Firstly, the experiment uses a limited number of comparative metrics, datasets, comparative MRs, and planning models, which may lack sufficient representativeness—though efforts have been made to mitigate this by selecting widely adopted metrics, general-purpose MRs, state-of-the-art planning models, and incorporating multiple datasets in each evaluation. Secondly, the evaluated task categories are restricted to navigation tasks, whereas real-world embodied agents engage in various other task types such as grasping, transporting, and cleaning. Additionally, violation detection relies on the AI2-THOR simulator to execute action sequences, but simulator-specific behaviors (e.g., simplified physics, idealized perception) may differ from real physical environments; discrepancies between simulated and real-world dynamics could lead to spurious NoD detections (e.g., a plan optimal in the simulator but suboptimal in reality).

Inherent limitations within the method itself also pose challenges to validity. This paper employs only four MRs, which cover merely a subset of the multiple properties exhibited in embodied agent task planning. These MRs, designed based on general optimality properties, may have limited coverage of task-specific optimality constraints—for instance, in healthcare assistance tasks, "safety" may be a more critical optimality criterion than "step efficiency", yet the current MRs do not explicitly encode such domain-specific properties, potentially leading to missed NoDs in specialized scenarios. Moreover, the optimality checker’s comparison mechanism assumes that the cost function (e.g., step count) fully reflects decision quality, but this oversimplifies real-world complexity. In practice, optimality may depend on multiple conflicting criteria (e.g., time, energy consumption, and error tolerance), and reducing it to a single metric could mask trade-off-related NoDs; for example, a plan with fewer steps but higher energy consumption might be deemed optimal by the current checker, even if it violates broader optimality principles.
\section{Related Work}
\label{related-work-sec}

\textbf{Embodied Agent Evaluation.} 
Existing research on embodied agent verification primarily focuses on functional correctness and domain specifications. Formal verification methods such as model checking mainly target safety-critical properties but rarely address non-functional attributes like plan cost or execution efficiency ~\cite{bensalem2014verification}\cite{orlandini2014planning}\cite{li2014model}. Moreover, existing evaluation work is severely insufficient when it comes to non-functional requirements, particularly the agents’ ability to make optimal decisions and generate optimal paths ~\cite{eniser2022metamorphic}. Current methods for testing AI planners primarily focus on two aspects: assessing the extent to which these planners comply with the Planning Domain Definition Language (PDDL) \cite{percassi2022power} ~\cite{smirnov2024generating}, and validating their input domains and problem definitions~\cite{magnaguagno2020web}. However, despite ongoing efforts to formally verify the correctness of generated plans~\cite{farias2017predicting}and test domain/problem definitions~\cite{magnaguagno2020web}, a significant challenge remains in determining whether the generated plans are optimal. Currently, there is no systematic evaluation method for the optimality of embodied agents~\cite{2024Decictor}. Existing research has not solved core problems such as ``how to define optimality", ``how to generate scenarios that can expose non-optimal decisions", and ``how to verify whether a decision is optimal".

\textbf{Optimality in Embodied Agents.} 
While some work pursues specific forms of optimality (e.g., shortest trajectories in path planning), systematic evaluation frameworks remain lacking. MT has been applied to action policies in gaming and planning contexts~\cite{eniser2022metamorphic,ayerdi2022performance}, employing state relaxation techniques to formulate MRs. However, these applications lack specific focus on optimality evaluation in embodied agents. Zhang et al.~\cite{zhang2019testing} utilized triangle inequality-based MRs to test graph-based path-planning algorithms, but this approach is limited to specific path-planning scenarios and lacks generalizability to broader embodied agent domains. Cheng et al.~\cite{2024Decictor} employed MT to evaluate learned policies in autonomous driving, where input scenarios are mutated to preserve optimality properties. While sharing conceptual similarities with our approach in MR design, their work focuses on policy evaluation in constrained road driving scenarios rather than general NoD detection across diverse embodied agent tasks. Mazouni et al.~\cite{mazouni2025mutation} integrated mutation testing with MT for AI planner optimality, introducing a mutation adequacy-based state selection strategy. However, existing approaches remain limited to specific domains and lack the diversity-guided test case selection mechanism necessary for comprehensive NoD detection across different violation types.

Our work addresses these limitations by proposing a systematic framework for NoD detection in embodied agents through diversity-guided metamorphic testing. Unlike prior work that focuses on domain-specific optimality or policy evaluation, we provide a comprehensive approach that combines carefully designed metamorphic relations with diversity-guided selection to detect diverse types of planning inefficiencies across different embodied agent architectures.

\section{Conclusion}
\label{conclusion-sec}

In this paper, we propose \method, the first metamorphic testing framework designed to evaluate the decision optimality of embodied agents. We formalize the Non-optimal Decisions (NoDs) problem and define four specialized metamorphic relations capturing fundamental optimality properties. Our approach systematically detects optimality violations through diversity-guided metamorphic transformations without requiring ground-truth optimal solutions, enabling comprehensive assessment of decision optimality across diverse planning scenarios.
The experimental results demonstrate the effectiveness and efficiency of NoD-DGMT across multiple planners and simulation environments. Our work addresses a critical gap in embodied AI evaluation and provides a systematic methodology that will guide the development of more efficient and reliable embodied agents for future real-world applications. 


\bibliographystyle{IEEEtran}

\bibliography{ref}

\end{document}